\documentclass[a4paper]{spie}  
\usepackage[english]{babel}
\usepackage{graphicx, subfigure}
\usepackage{float, epsfig, gensymb}
\usepackage{pstricks}
\usepackage{bbding} 
\usepackage{layout}
\usepackage{mathptmx} 
\usepackage{amssymb,amsmath,MnSymbol}
\usepackage{bbding}
\usepackage{keystroke}
\usepackage{pifont}        
\usepackage{subfigure}
\usepackage{float}
\usepackage{wrapfig}
\usepackage{tabularx}
\usepackage[latin1]{inputenc}
\usepackage{times}
\usepackage[T1]{fontenc}

\newcommand{\Beatrix}{BEaTriX }

\title{BEaTriX, expanded X-ray beam facility for testing modular elements of telescope optics: an update}
\author{C. Pelliciari\supit{1}, D. Spiga\supit{1}, E. Bonnini\supit{2}, E. Buffagni\supit{2}, C. Ferrari\supit{2}, G. Pareschi\supit{1}, G. Tagliaferri\supit{1}
\skiplinehalf
\supit{1} INAF / Brera Astronomical Observatory, Via Bianchi 46, 23807 Merate, Italy\\
\supit{2} CNR-IMEM, Parco Area delle Scienze 37/A, 43124 Parma, Italy
}
\authorinfo{Contact author: Carlo Pelliciari, e-mail: carlo.pelliciari@brera.inaf.it, phone: +39-02-72320472}

\begin{document} 
\maketitle 

\begin{abstract}
We present in this paper an update on the design of BEaTriX (Beam Expander Testing X-ray facility), an X-ray apparatus to be realized at INAF/OAB and that will generate an expanded, uniform and parallel beam of soft X-rays. BEaTriX will be used to perform the functional tests of X-ray focusing modules of large X-ray optics such as those for the ATHENA X-ray observatory, using the Silicon Pore Optics (SPO) as a baseline technology, and Slumped Glass Optics (SGO) as a possible alternative. Performing the tests in X-rays provides the advantage of an in-situ, at-wavelength quality control of the optical modules produced in series by the industry, performing a selection of the modules with the best angular resolution, and, in the case of SPOs, there is also the interesting possibility to align the parabolic and the hyperbolic stacks directly under X-rays, to minimize the aberrations. However, a parallel beam with divergence below 2~arcsec is necessary in order to measure mirror elements that are expected to reach an angular resolution of about 4 arcsec, since the ATHENA requirement for the entire telescope is 5~arcsec. Such a low divergence over the typical aperture of modular optics would require an X-ray source to be located in a several kilometers long vacuum tube. In contrast, BEaTriX will be compact enough (5 m x 14 m) to be housed in a small laboratory, will produce an expanded X-ray beam 60~mm x 200~mm broad, characterized by a very low divergence (1.5 arcsec HEW), strong polarization, high uniformity, and X-ray energy selectable between 1.5~keV and 4.5~keV. In this work we describe the BEaTriX layout and show a performance simulation for the X-ray energy of 4.5~keV.
\end{abstract}


\section{INTRODUCTION}\label{sec:introduzione} 

X-ray observatories of the future will be characterized by very large collecting areas and focal lengths of tens of meters, such as ATHENA. With a 3 m diameter optical module\cite{Athena-spie2014}, a 12 m focal length, and a required effective area of 2~m$^2$ at 1~keV, ATHENA will be the largest X-ray observatory ever built. However, in order to get high sensitivity and avoid source confusion, in addition to the large effective area there is a tight requirement on the angular resolution of 5~arcsec HEW (Half Energy Width), with a 3~arcsec goal. The fabrication of monolithic X-ray mirrors of such dimension is not realistic: hence, the ATHENA optical module will consist of nearly 1000 modular X-ray Optical Units (XOU), each of them made by stacking some tens of parabolic/hyperbolic lightweight segments with high focusing performance, and carefully aligned to reconstruct the Wolter-I configuration\cite{wolter-optics}. Finally, these modular elements are carefully aligned into a supporting structure in order to attain the required 5~arcsec HEW. 

Lightweight modular optics are being developed at ESA/ESTEC and Cosine\cite{poreoptics-esa2013} since 2004 on the basis of the SPO (Silicon Pore Optics) technology\cite{poreoptics-2012}, currently the baseline for the ATHENA optics. SGO (Slumped Glass Optics) are also developed in parallel at INAF/OAB\cite{Salmaso2014} and MPE\cite{Proserpio2015} (Max Planck Institute for Extraterrestrial Physics, Garching) as a possible backup technology. However, in both cases a very high number of XOUs ($>$1000 for the SPO case) have to be produced in an industrial environment, and accounting for unavoidable alignment errors, the individual XOUs must have an individual HEW that is better than 5~arcsec. For this reason, in-situ quality controls have to be routinely performed to discard/replace the XOUs with insufficient angular resolution.  

In addition to standard metrology tests (mirror plate profile and roughness), a selection criterion can be a direct measurement of the angular resolution in X-rays. In fact, usual metrology tools can hardly inspect the optical profile once the ribbed plates are densely stacked, and tests with optical or UV light are not only dominated by aperture diffraction (that is usually negligible in X-rays), but are also insensitive to profile undulations with period in the centimeter range or smaller, not to mention the surface roughness. X-ray test are, in contrast, fully representative of the at-wavelength performances of the optical module. An obvious difficulty comes, indeed, from the need to simulate the X-ray flux from an astronomical source, i.e., broader than the XOU aperture, uniform, and parallel -- or, at least, much less divergent than the XOU HEW to be measured. This cannot be simply obtained by locating a commercial X-ray source at a few meters distance from the XOU aperture (as it would desirable to fit the facility within a small size): for example, the detector-to-source distance has to be $S > 4f$, where $f$ is the focal length of the XOU in order to have a focus at all. Other effects related to the finite distance of the source\cite{spiga-spie2014} can be mitigated by a proper adjustment of the incidence angle, but the wavefront curvature cannot be removed easily and results into a blur of the focal spot, which is difficult to disentangle from the PSF of the XOU under test. Finally, for SPOs there is also the problem of a severe vignetting caused by the ribs if the beam is divergent.

Currently, XOU modules are characterized in X-rays with a pencil beam at PTB laboratory of BESSY synchrotron facility\cite{pencil-spie2013}) and with a low-divergent beam at PANTER facility\cite{XOU-Panter-spie2013}. In the first case, the surface is probed pore-by-pore by a narrow beam by dithering the XOU under the beam: the Point Spread Function (PSF) is therefore to be reconstructed superposing the images and accounting for the exact displacement of the XOU during the X-ray exposure. Performing functional tests of all the focusing elements produced can be not so easy.

At PANTER, the divergent beam is generated by an X-ray source located at a 123~m distance, reducing the beam divergence to a 1.6~arcsec/mm with respect to the central axis. Work is in progress at PANTER to remove this divergence by means of transmission zone plates\cite{Menz2014}, but also in this case the source has still to be located at a very large distance, implying large dimensions and volumes of the facility. PANTER can be used to test large XOU assemblies (e.g. petals) of ATHENA, but hardly to routinely perform the functional tests of nearly 1000 XOUs. Moreover, these X-ray facilities cannot be moved or reproduced at the production/qualification site. 

\begin{figure}[ht]
	\centering
	\includegraphics[width=0.8\linewidth]{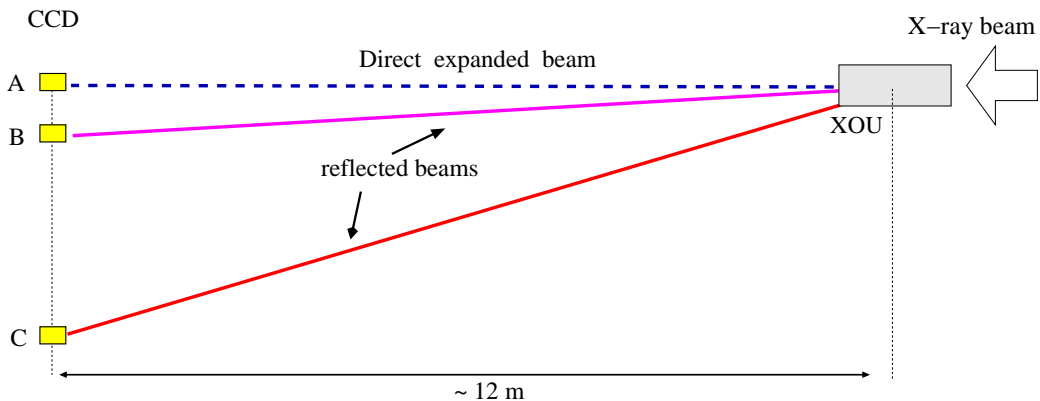}
	\caption{XOU Characterization concept at the \Beatrix facility: the focused beam is deviated downwards by the optical element under test.}
	\label{fig:lens-beam}
\end{figure}

With the specific aim of performing the functional tests of the ATHENA XOU, we are designing the \Beatrix X-ray facility\cite{spiga-spie2014, spiga-spie2012}, a compact X-ray system (5 m x 14 m) to be installed at INAF/OAB, and subsequently replicated at the industrial production site for performing the functional tests of the optic segments of a large X-ray telescope like ATHENA. The system will produce a broad (200 mm x 60 mm), parallel (divergence < 1.5 arcsec HEW), and uniform beam of monochromatic X-rays selectable between 1.5 keV and 4.5 keV. The beam is strongly polarized in the incidence plane of the optic under test, enabling tests of polarimetric X-ray instrumentation. The entire system fits a small-size laboratory, and the reduced volume will also make the venting/sample change/pumping down a few hours matter. In this way, a larger number of samples can be tested in a given time, and a quicker feedback after the manufacturing can be provided.

If the expanded beam is used to illuminate the aperture of a XOU, the X-ray beam is focused at its nominal focal distance and deviated downwards (Fig.~\ref{fig:lens-beam}) by an angle $4\alpha$, where $\alpha$ is the incidence angle for an on-axis beam. The focal spot is directly seen in the recorded image on a camera, returning a direct measurement of the PSF and -- by comparison with the direct beam -- the effective area. The characterization will be in situ and in real time. Finally, in the case of XOUs based on the SPO technology, there will also be the interesting possibility to align the parabolic to the hyperbolic stacks: the best alignment will be achieved when the effective area is maximized and the HEW is minimized under X-ray illumination.

In the next sections we provide an updated layout of the \Beatrix design, describe the components used to expand and collimate the beam, and finally provide a simulation of the system using the SHADOW computer code, verifying the expected optical performances. 

\section{Facility Design \label{sec:metodo}}
The method that will be implement is an extension of the concept already implemented successfully in 1994 at the Daresbury synchrotron\cite{Finn1994daresbury}, consisting in a wavefront expansion via diffraction onto an asymmetric cut crystal. In that case, however, the equipment was expanding the beam in only one direction and the divergence was reduced to 20~arcsec, which is too much for our scopes. In contrast, we want to expand the beam in two dimensions to a size of 20 cm $\times$ 6 cm. This beam size has been selected in order to illuminate the XOUs for ATHENA up to the largest ones. More exactly, the 20~cm width is determined by the maximum size of Silicon wafers available commercially, while 6~cm of height have been conservatively assumed to allow us to test also XOUs based on SGOs. Finally, the expansion should preserve the beam collimation, which should be better than 1.5~arcsec HEW.

\begin{figure}[ht]
	\centering
	\includegraphics[width=0.8\linewidth]{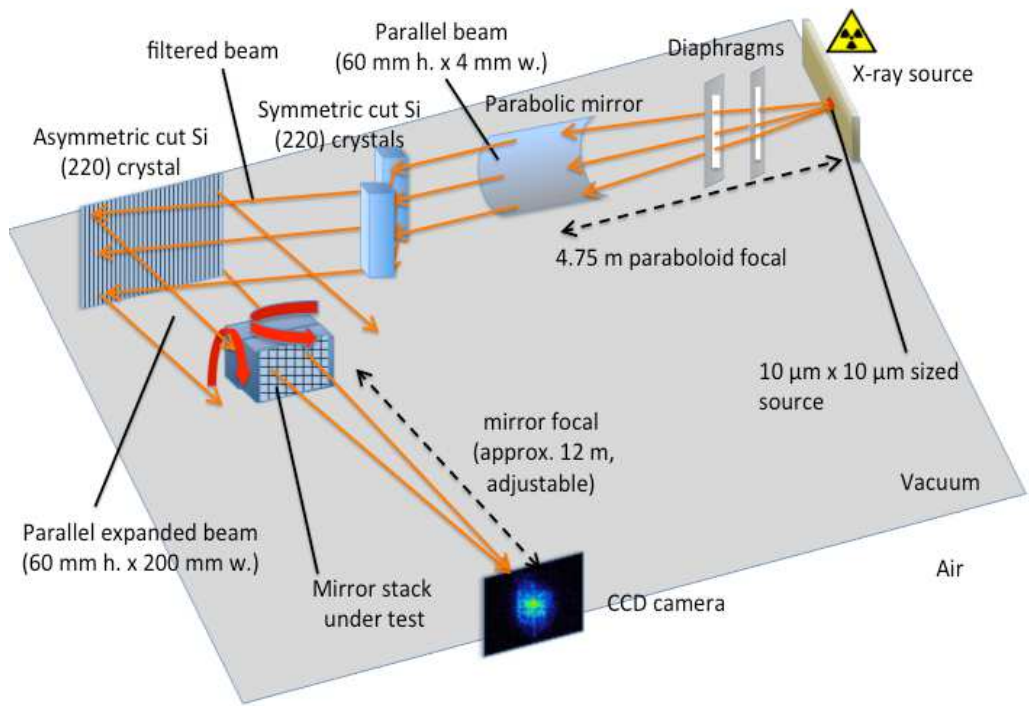}
	\caption{Optical layout for the 4.5~keV X-ray energy. After the beam collimation by the paraboloidal mirror, two symmetric Silicon crystals are used to filter the fluorescence line. The subsequent beam expansion is demanded to a Silicon crystal with asymmetric cut. The configuration for the 1.5~keV energy should make use of ADP crystals.}
	\label{fig:layout}
\end{figure}

To obtain a bidimensional beam expansion, a paraboloidal, grazing-incidence mirror has been located 5~m downstream the X-ray source, which in this way occupies its focus (Fig.~\ref{fig:layout}). The scope of the mirror is twofold: it collimates the beam, making the wavefront planar after the reflection, and expands the beam in the vertical direction. The beam collimation also enables a constant incidence angle on the diffracting crystals and a uniform diffraction efficiency. 

\begin{figure}[ht]
	\centering
	\includegraphics[width=0.78\linewidth]{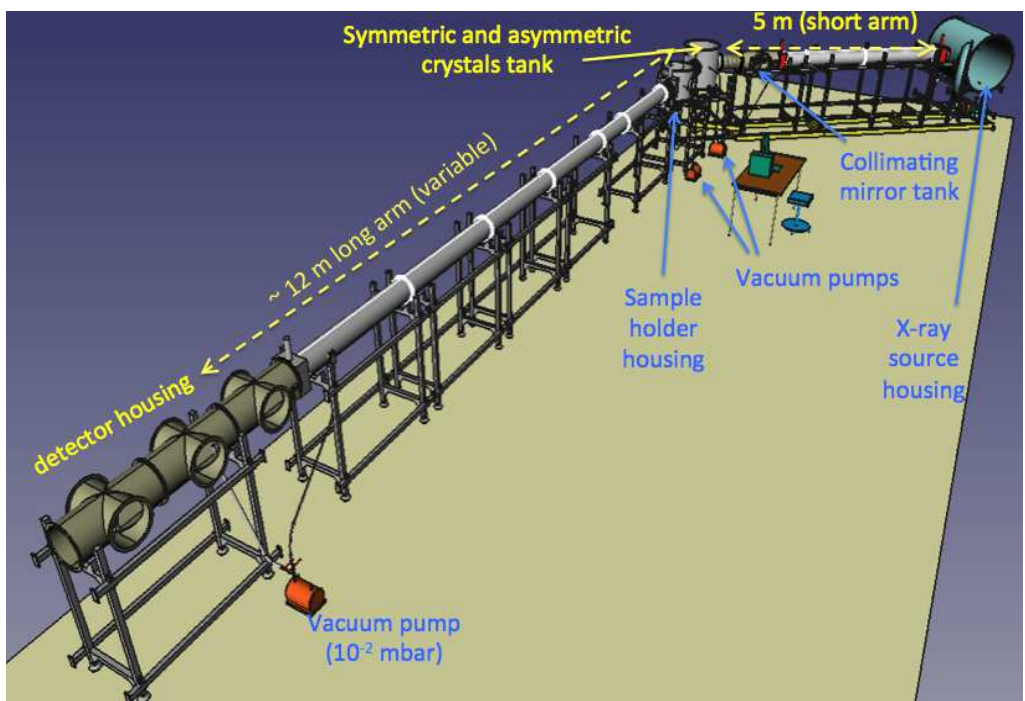}
	\caption{Outer view of the \Beatrix facility design. The short arm in the background is used to obtain the beam collimation and expansion. The beam reflected by the XOU under test propagates to focus in the long arm in the foreground. The system should be kept at steady temperature (within 1~$^{\circ}$C) to precisely maintain the alignment of the components.}
	\label{fig:outer_layout}
\end{figure}

The feasibility of \Beatrix has been already demonstrated\cite{spiga-spie2014}. The performance of the collimated beam in terms of collimation, intensity, and uniformity, crucially depends on the optical quality of the components. For example, the X-ray source has to be very small (less than 30~$\mu$m required, with a 10~$\mu$m goal) in order to produce a accurately spherical wavefront and, after the collimation, maintain the residual divergence below 1~arcsec in the vertical plane. Profile and surface of the collimating mirror should also be accurately figured to a few arcsec HEW in order to avoid a loss of diffracting efficiency in the monochromators (the divergence in the horizontal plane is not affected by the mirror figure accuracy to within large limits\cite{spiga-spie2014}). All the system is contained within vacuum tubes (Fig.~\ref{fig:outer_layout}), and, owing to the short range of photons (17 m), a low vacuum (10$^{-2}$~mbar) is sufficient to avoid X-ray absorption and can be obtained with dry pumps in one hour evacuation time.

The facility is currently in design phase, while the component procurement and construction of the facility will enter the operative phase in September 2015, thanks to a grant awarded by the EU to the AHEAD collaboration, which includes also the \Beatrix realization. We describe the characteristics of the optical components in the next paragraphs.

\subsection{The X-ray source}\label{sec:source}
As mentioned above, the source should be very small to reduce the collimation as much as possible. The vertical size directly affects the vertical divergence of the beam, while the horizontal size causes a dispersion of the incidence angles on the crystals and a severe beam intensity reduction. To minimize optical aberrations, the X-ray source has to be aligned with an accuracy better than 35~$\mu$m in the plane perpendicular to the optical axis and 1~mm in the longitudinal direction.

Conventional bremsstrahlung sources are not suitable in general, because of their low efficiency (only 5\% of the energy is converted to X-ray). Moreover, the source of X-rays has sizes of 100~$\mu$m or more, which does not enable the required collimation. For this reason, we are considering the procurement of a micro-focus source, which enable copious brilliances (typical intensities are of 10$^{10}$ ph/sec/sterad), small source sizes (down to 5 $\mu$m) and very low power consumption (a few tens of watts). Two sources should be purchased, one with aluminum anode (1.49 keV, K$\alpha$ fluorescence line) and another one with titanium anode (4.51 keV, K$\alpha$ fluorescence line), both in the energy band of ATHENA (0.2-12 keV). Performing the characterization at two energies will enable us to disentangle X-ray scattering from the effect of figure errors.
	
\subsection{Apertures/slits}
The source is in the focus of the collimating mirror at a distance of about 5~m. A set of multiple slits has to be mounted to reduce the scattering and reducing the stray light in the apparatus. The apertures should have a sawtooth profile (single or double tooth) with the tilted surface facing the X-ray source, in order to avoid scattered photons off the lateral surface of the slits.

\subsection{Paraboloidal Mirror \label{sec:paraboloid}}	
\begin{figure}[ht]
		\centering
		\includegraphics[width=0.95\linewidth]{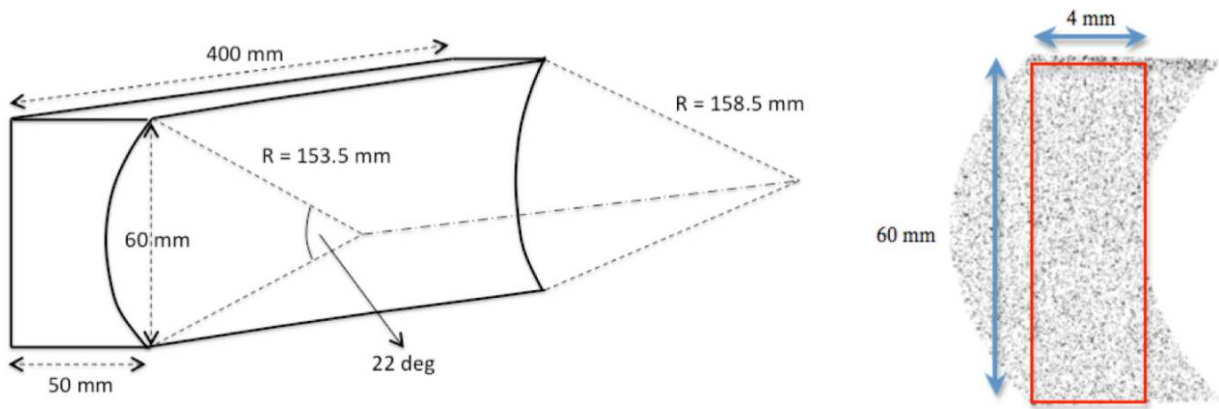}
		\caption{(left) a drawing of the paraboloidal mirror. (right) ray-tracing from the source, showing the ray positions at the exit pupil. The red square is the portion of the beam that is collected by the asymmetric crystal.}
		\label{fig:parabola-forma}
\end{figure}

The first optical component is a grazing incidence mirror shaped as a paraboloidal sector, with the X-ray source in its focus ($\sim$ 5~m), which parallelizes the beam. The focal length is 4.75 m and the grazing incidence angle is 0.93 deg. The beam section after reflection in the shape of a "crescent", in which a 60~mm $\times$ 4~mm rectangle can be inscribed (Fig.~\ref{fig:parabola-forma}). This rectangle represents the part of the beam that is collected and will subsequently be expanded horizontally by the asymmetric crystal (Sect.~\ref{sec:asym}), in order to illuminate the entrance pupil of the XOU under test. To ensure a proper performance of \Beatrix, the surface quality of the collimating mirror (HEW $< $ 5~arcsec) is very important. In fact, figure errors or a non-negligible surface roughness scatter the beam, mostly in the incidence plane: the incidence angle on the monochromators would then get spread, making the final beam non-uniform if the angular dispersion becomes comparable to the width of the rocking curve. The resulting angular spread in the incidence plane, however, does not affect directly the final collimation of the expanded beam, as we see in Sect.~\ref{sec:asym}, because in asymmetric configuration the beam divergence is de-magnified by a factor equal to the asymmetry factor. Moreover, the vertical collimation is almost unaffected, because in grazing incidence reflection sagittal defects have a much lesser weight than axial ones.

A suitable material to manufacture the mirror can be Zerodur\texttrademark, owing to its extremely low thermal expansion, and because it can be polished to an excellent level. To reduce machining time and costs, the collimating mirror can be figured from a purchased conical segment, using one of the two Ion Beam Figuring machines operated at INAF/OAB\cite{ghigo2014spie, ghigo2007spie}. The mirror polishing can be achieved with a lapping machine developed in our labs. Eventually, a 30~nm Platinum coating plus a 5~nm of amorphous Carbon\cite{cotroneo2008spie} endows the mirror with a reflectivity of 89\% at 1.49 keV and of 41\% at 4.51 keV.

\subsection{Symmetric crystals \label{sec:symmetric}}
As already mentioned, the beam expansion in the horizontal direction is demanded to an asymmetric diffraction. However, after the collimation by the parabolic mirror, the beam is still polychromatic (fluorescence lines+continuum), and the energy of interest must be filtered in order to ensure a final collimation of the beam. Computation shows that the spectral purity needed to remain with a divergence of 2~arcsec FWHM (approximately corresponding to a 1.5~arcsec HEW for a Gaussian distribution) is near the natural width of the X-ray line (0.2~eV). In absence of spectral filtering, the beam would exhibit a divergence of about 10~arcsec, which is clearly unacceptable.

\begin{figure}[ht]
	\centering
	\subfigure[]{\label{fig:Si111-Si220}\fbox{\includegraphics[height=6.1cm]{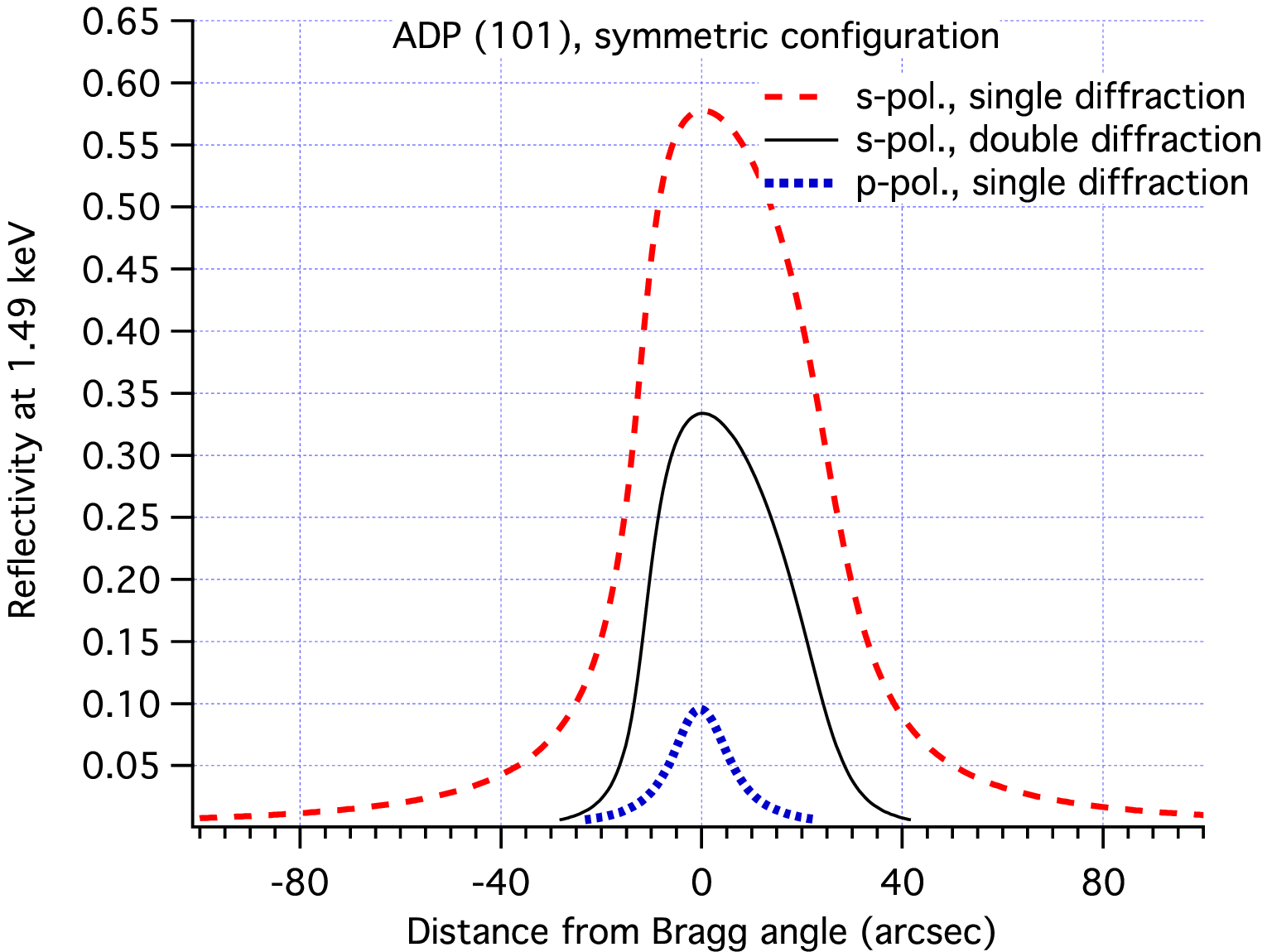}}}
	\subfigure[]{\label{fig:Si111-Si220-4diff}\fbox{\includegraphics[height=6.1cm]{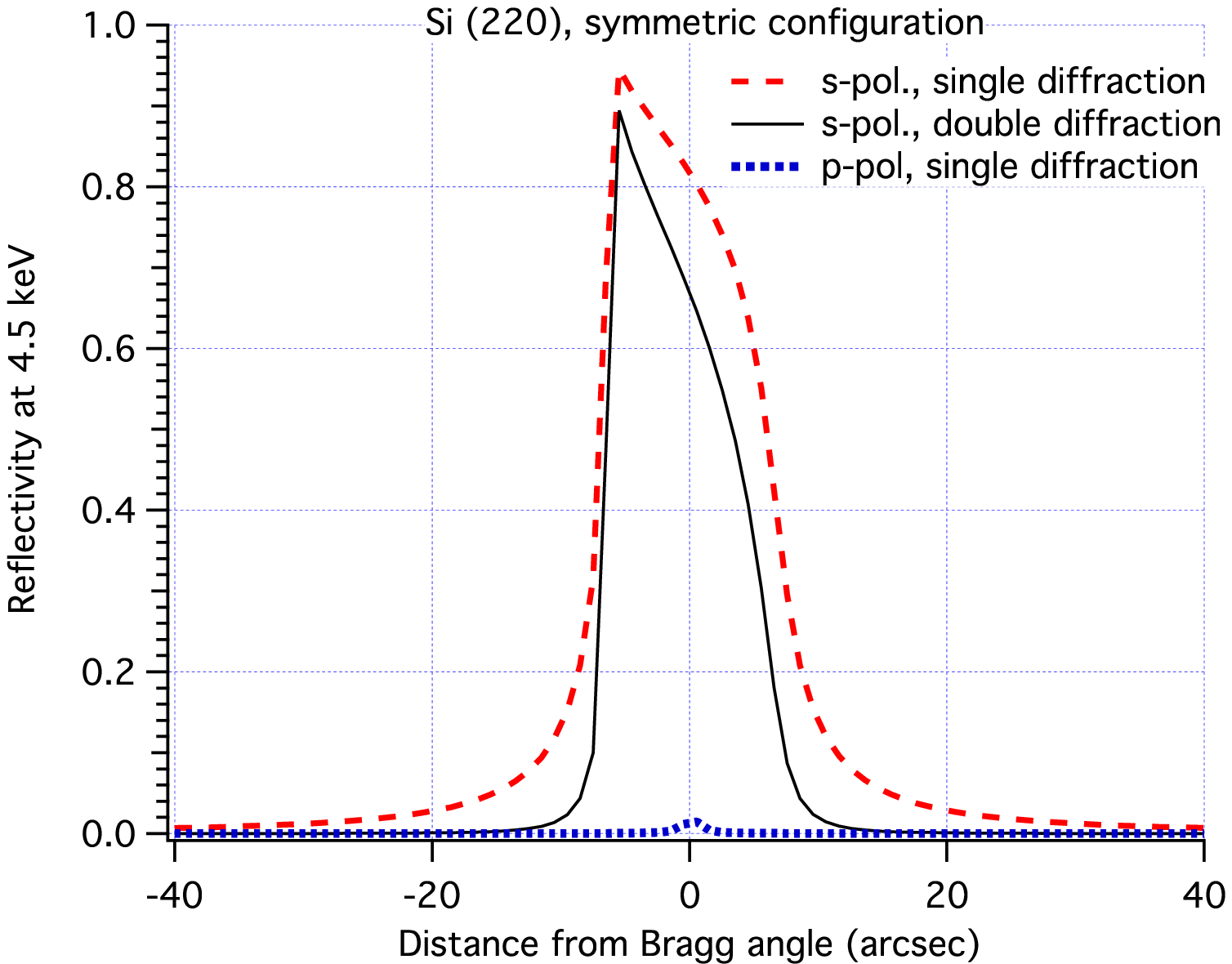}}}
	\caption{a) Symmetric diffraction on ADP(101). b) Symmetric diffraction on perfect Si(220). For Silicon, a repeated diffraction improves the monochromation. In both cases, the diffraction of the s-polarization is much more intense than the p-polarization.}
	\label{fig:Si111-Si220-diff}
\end{figure}

Therefore, a tight monochromation is necessary: it is obtained via a diffraction off two or more crystals with symmetrical cut, i.e., with crystalline planes parallel to the incidence surface. Assuming a flawless crystalline structure, in symmetric configuration the diffracted beam has the same size of the incident one; moreover, if the incidence angle is precisely set at the Bragg angle for the energy of the fluorescence line, the continuum is discarded and the X-ray line is reflected. 

For the 4.5 keV setup, the (220) Silicon crystal diffraction can be used, corresponding to a d-spacing $d$ =1.91~\AA~and a Bragg angle $\theta_{\mathrm B}=$45.847~deg off-surface. At 1.5~keV, $\lambda > 2d$ and no diffraction is possible with Silicon: at this energy, organic crystals such as ADP (Ammonium Dihydrogen Phosphate, NH$_4$H$_2$PO$_4$) should be used. The rocking curves are shown in Fig.~\ref{fig:Si111-Si220-diff}. ADP crystals with the required plane orientation are commercially available.
	
\begin{figure}[bh]
	\centering
	\includegraphics[width=0.55\linewidth]{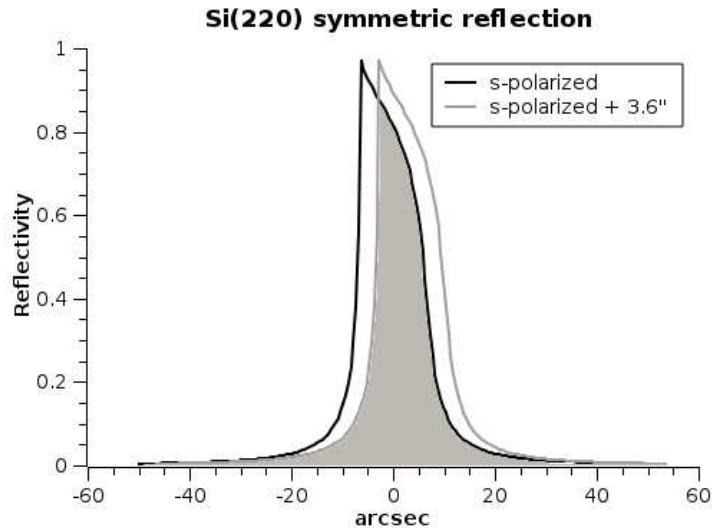}
	\caption{Rocking curve superposition to shrink the bandpass of symmetric Si (220) at 4.5 keV, obtained by misaligning two nearly-parallel crystals (by 3.6~arcsec).}
	\label{fig:Si220-diff}
\end{figure}

Because the rocking curves of crystals have a finite width, the monochromation level can be insufficient. For Si (220), an angular FWHM of approx. 15~arcsec corresponds to a $\Delta E = E\cdot\Delta\theta_{\mathrm B}\cdot\mathrm{cotan}\theta_{\mathrm B}$ = 0.3~eV, which is close to the natural line width. However, the rocking curves still shows pronounced tails that shall be removed. To improve the monochromation, two symmetric diffractions can be used (Fig.~\ref{fig:Si111-Si220-diff}, b). The (220) plane orientation has been selected because of the particular profile of the rocking curve, which exhibits a sharp peak with a reflectivity of 95\%. In this way, the peak reflectivity remains high, while the tails are suppressed. Another advantage of a double diffraction on parallel planes is that the direction of the beam remains unchanged, reducing the number of movable mechanical parts (e.g., the joint between the short and the long arm described in Sect.~\ref{sec:arm1} and~\ref{sec:arm2}) and avoiding an excessive beam folding. Should a further monochromation improvement be necessary, the two crystals can be slightly misaligned (Fig.~\ref{fig:Si220-diff}). In this way, the angular scales of the two rocking curves is shifted and the spectral response becomes much narrower. 

Extension to a higher number of diffractions for a tighter monochromation is possible, but it is always convenient to use an even number of crystals to keep the beam direction unchanged. For example, one or more Channel Cut Crystal (CCC) configurations might be used, with the additional possibility to tilt one of the pairs. The CCC orientation can be controlled by stepper motors.

Finally, especially for Silicon, the incidence angle is very close to the polarization angle. As a result, the expanded beam will be almost completely polarized (Fig.~\ref{fig:Si111-Si220-diff}) in the vertical plane. For ADP, the polarization is less pronounced, but still high (on the order of 80\%). The use of a polarized beam does not affect the PSF and the effective area of the XOU, because in grazing incidence the reflectivity is almost exactly same for both polarization states, but it provides the additional opportunity to test X-ray polarimetric systems\cite{Fabiani}. 

\subsection{Asymmetric Crystal \label{sec:asym}}
\begin{figure}[ht]
		\centering
		\includegraphics[width=0.85\linewidth]{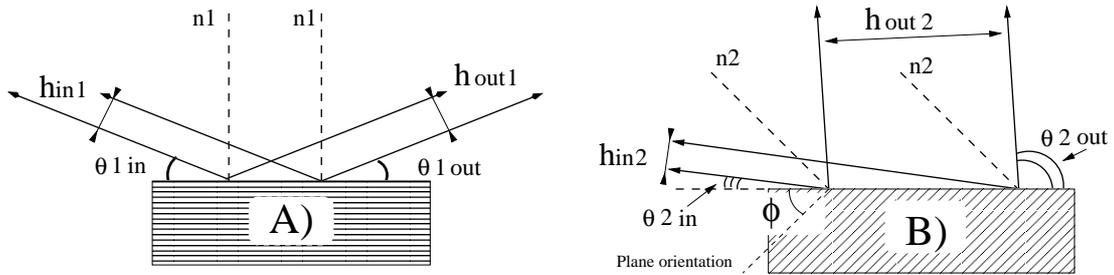}		
		\caption{Diffraction off a crystal with symmetric (A) and asymmetric cut (B). In the asymmetric configuration, the beam is expanded by the diffraction, because it behaves like a reflection occurring with respect to the crystalline planes and not to the outer surface.}
		\label{fig:cristalli}
\end{figure}

The filtered and collimated beam finally impinges onto	the asymmetric crystal that achieves the horizontal expansion to a 200~mm size. The crystal is illuminated in grazing incidence (at an angle larger than the critical one for total external reflection), but the beam diffraction occurs with respect to the crystalline planes, which in asymmetrical cut form an angle $\phi$ with the outer surface. If the asymmetry angle is properly chosen, the beam fulfills the Bragg condition and is diffracted at the angle $\theta_{\mathrm B}+\phi$ off-surface. Because the incidence angle on the surface is $\theta_{\mathrm B}-\phi$ (Fig.~\ref{fig:cristalli}), the beam is expanded by the asymmetry factor\cite{zach, james}:
\begin{equation}
	b = \frac{\sin(\theta_{\mathrm B} + \phi)}{\sin(\theta_{\mathrm B} - \phi)}.
	\label{eq:F}
\end{equation}
Figure~\ref{fig:cristalli} shows an example of reflection of the same beam in symmetric (A) and asymmetric configuration (B). Replacing the numbers for Si(220) and 4.5 keV, $\theta_{\mathrm B}$ = 45.847~deg, $\phi$ = 44.8~deg, we obtain $b \approx 54$, i.e., the required expansion factor. The length of the crystal needed is $h/\sin{(\theta_{\mathrm B} - \phi)}$, where $h$ = 4~mm is the beam section before the expansion. The result is confirmed by simulation using the SHADOW software package\cite{XOP2004}.		
\begin{figure}[ht]
		\centering
				\includegraphics[width=0.7\linewidth]{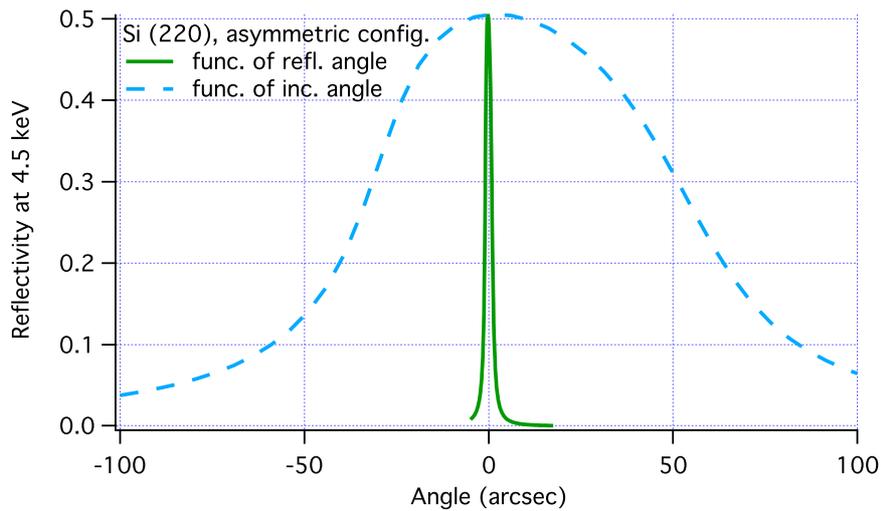}		
		\caption{Diffraction curve of the asymmetrical Si(220). The reflectivity is displayed as a function of the incidence angle and the reflection angle. Because of the asymmetric configuration, the acceptance angle in entrance is much broader than the possible range of the exit angles.}
		\label{fig:asymm_conf}
\end{figure}

In addition to the beam section expansion, an interesting aspect of the asymmetric diffraction is the asymmetric behavior in the angular dispersion. The acceptance in the incidence angles, $\Delta \theta_{\mathrm{in}}$, is $b$ times larger than the angular distribution of diffracted radiation, $\Delta \theta_{\mathrm{out}}$, in agreement with Liouville's theorem (in 1 dimension, refer to Fig.~\ref{fig:cristalli} for symbol explanation):
\begin{equation}
	\Delta\theta_{\mathrm{in}}\,h_{\mathrm{in}} \approx \Delta\theta_{\mathrm{out}}\,h_{\mathrm{out}}.
	\label{eq:liou}
\end{equation}
For example, in Fig.~\ref{fig:asymm_conf} we show the diffraction curve for Si(220) as a function of the incidence and the reflection angle: while in incidence the FWHM of the curve is 100~arcsec, in reflection it is 54 times smaller, i.e. 2~arcsec. This means that the beam is expanded, but also {\it collimated} by the asymmetric diffraction. Owing to the large acceptance angle, the beam does not forcedly need to be tightly collimated before the asymmetric diffraction; hence, the parabolic mirror does not need to be polished to sub-arcsec accuracy. The horizontal HEW after the diffraction is 2~arcsec, corresponding to 1.5~arcsec if the curve can be approximated by a Gaussian. As already mentioned, the vertical HEW is mostly determined by the size of the X-ray source.
		
\subsection{XOU manipulator and detector} 
After the beam expansion, the parallel beam is used to illuminate the aperture of the XOU under test. The alignment is achieved using a manipulator with precise stepper motors. The beam is polarized in the incidence plane of the XOU. 
After the reflection, the focused beam propagates in the 12 m-long vacuum tube, which can be steered to fit the incidence angles on the optical element under test. The PSF is recorded in the focal plane. To ease the measurement, the X-rays are converted in visible light by a thin phosphor screen and then imaged by a CCD outside the vacuum tube. Devices of this kind are commercially available. Intra- or extra-focus measurements are possible removing or adding tube segments and changing the XOU-to-camera distance.
	
\subsection{Mechanical Layout \label{sec:mechanical}}
\begin{figure}[b]
\centering
	\includegraphics[width=0.80\linewidth]{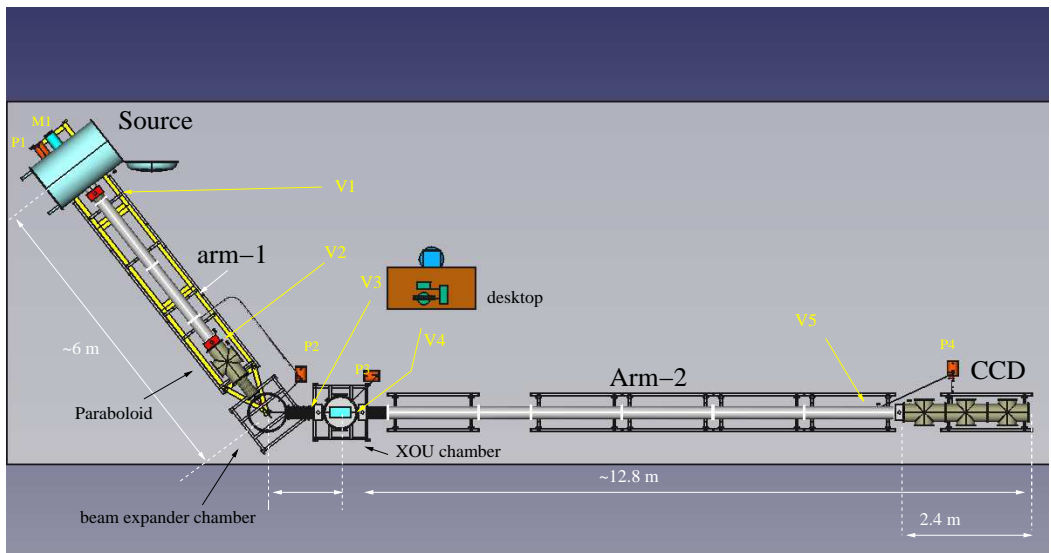}
	\caption{\Beatrix mechanical layout, top view. V1-5: valves isolate section of tube in order to reduce the venting/XOU mounting/pumping down time. P1-4: pumps. M1: system for the angular movement of the short arm.}
	\label{fig:beatrix-top-en}
\end{figure}

A mechanical layout is reported in Fig.~\ref{fig:beatrix-top-en}. The facility is about to be built in the basement laboratories at INAF/OAB, Merate. The facility is composed by two main parts: a "short" arm (approx. 5~m long) in which the optical components dedicated to the production of the expanded beam and a "long" arm (approx. 12~m long) to house the XOU to be tested and the propagation of the focused beam to the detector plane. 

\subsubsection{Short arm, production of the parallel beam \label{sec:arm1}}
\begin{figure} [th]
	\centering
	\includegraphics[width=0.85\linewidth]{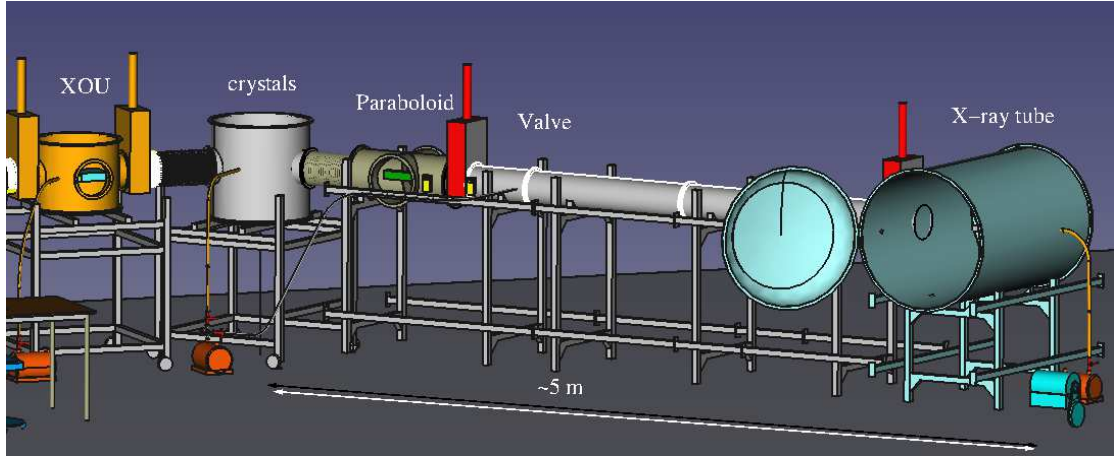}
	\caption{Detailed layout of the short arm of \Beatrix.}
	\label{fig:arm1}
\end{figure}

A detailed view of the short arm is reported in Fig.~\ref{fig:arm1}. The short arm can be steered in the horizontal plane about an axis passing by the center of the asymmetric crystal via sealed, flexible joints in order to align the vacuum tube with the X-ray beam. The reason is that the alignment changes passing from the 1.5~keV to the 4.5~keV setup because of the different incidence angle on the asymmetric crystal. To this aim, a flexible metal bellows links the crystal chamber with the short arm, enabling the angle change within a 20-30 deg range. The vacuum tube is sectioned to allow us venting only the parts that need to be opened for operation.
		
The X-ray tube chamber is designed to contain the required X-ray tubes, one for each energy line at which the XOU characterization will be performed. All X-ray tubes will be mounted on linear stages that can move the X-ray source of interest in front of the first collimating slit without breaking the vacuum. The linear stage has to set the position of the source with accuracy better than 35 $\mu$m in the plane orthogonal to the X-ray beam. The paraboloidal mirror will be placed at about 5~m from the source, in a tank with a lateral aperture to ease the access for assembly and maintenance. The paraboloid positioning will be ensured by a linear and rotational stage.
		
\begin{figure}[hb]
	\centering
	\subfigure[]{\label{fig:arm2A}\includegraphics[height=0.3\linewidth]{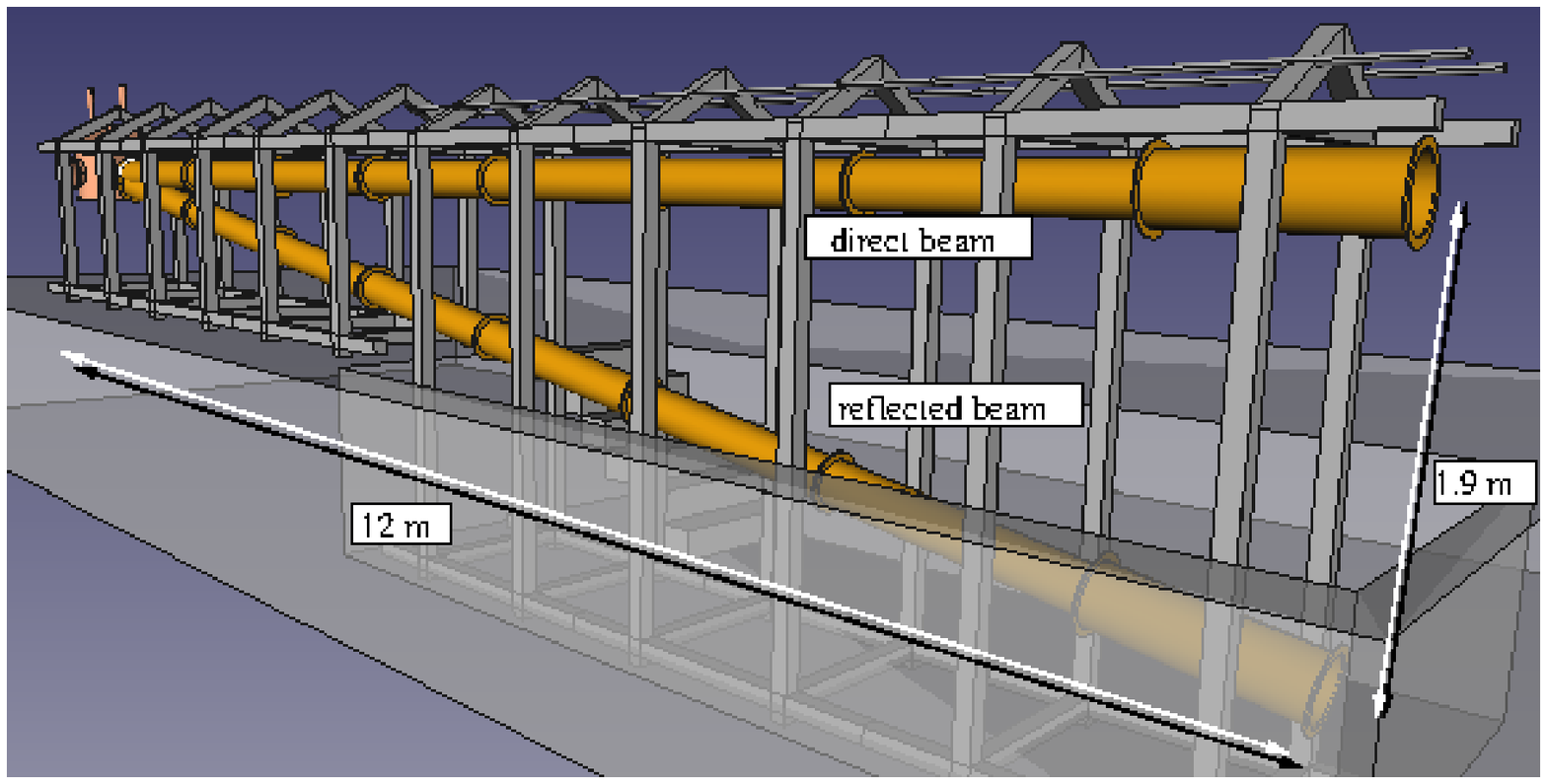}}
	\hspace{0.02\linewidth}
	\subfigure[]{\label{fig:arm2B}\includegraphics[height=0.3\linewidth]{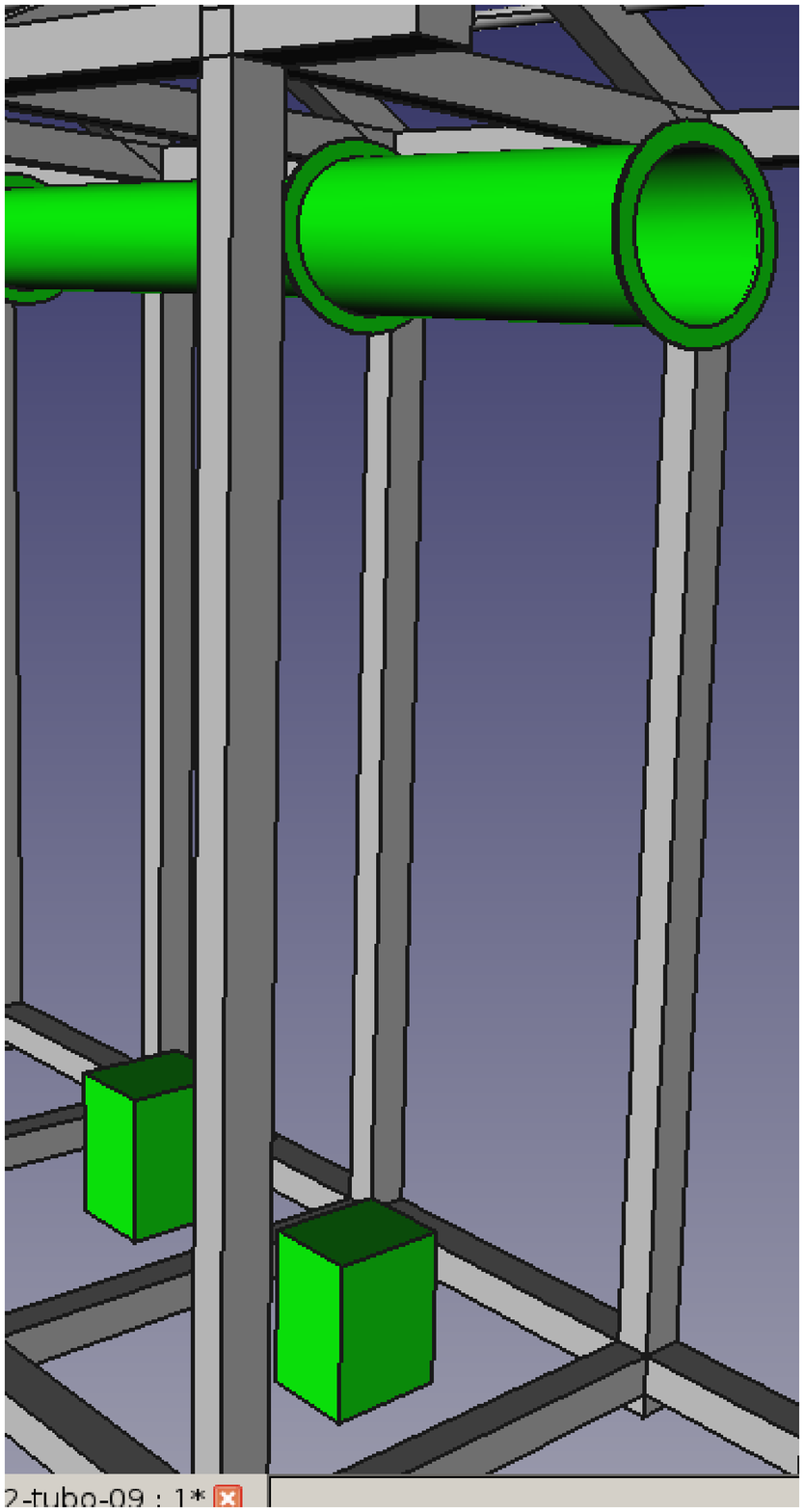}}
	\hspace{0.02\linewidth}
	\subfigure[]{\label{fig:arm2C}\includegraphics[height=0.3\linewidth]{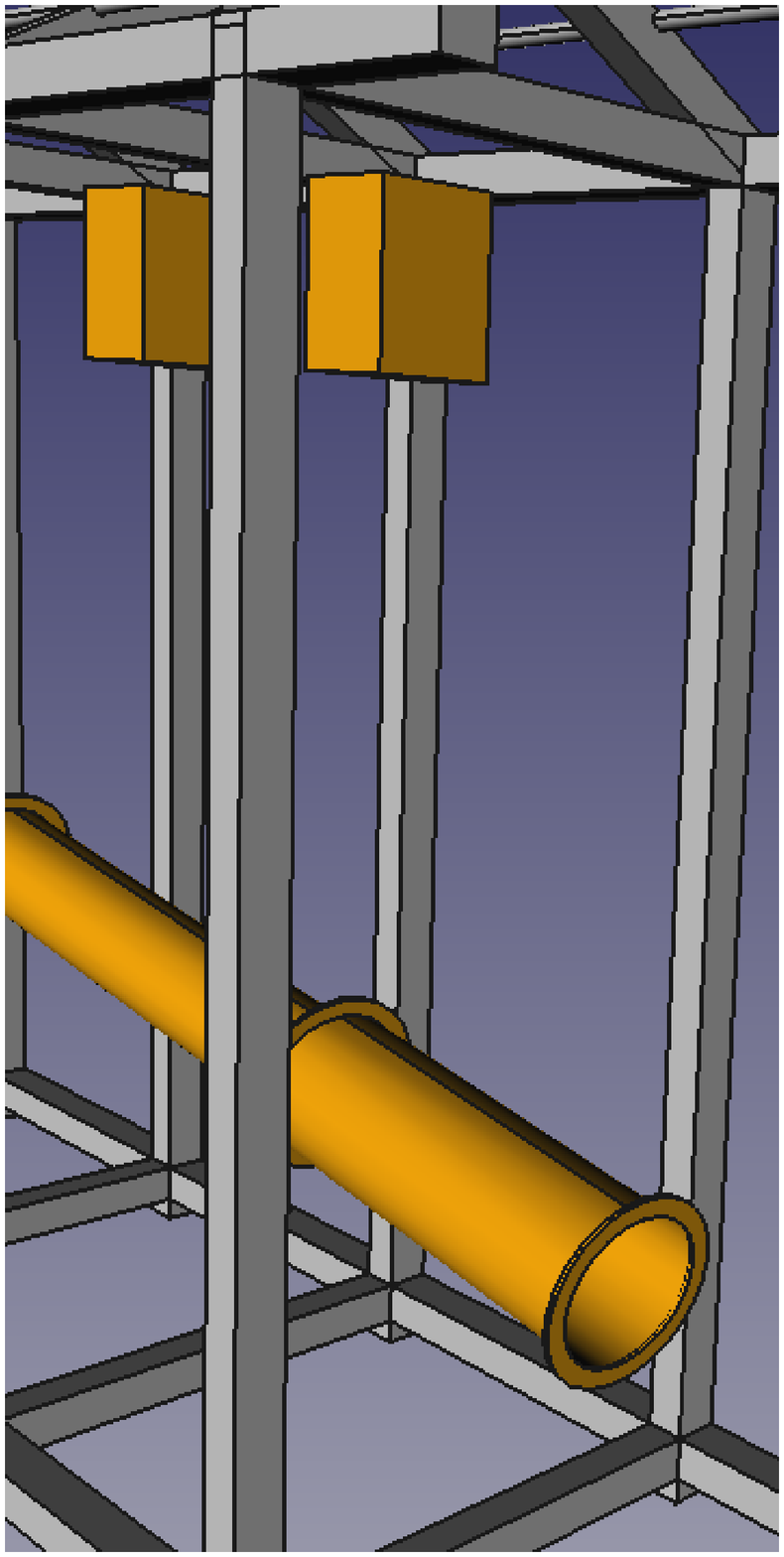}}
	\caption{a) Layout of the long arm of \Beatrix. The pipeline can tilt (b) up and (c) down to follow the beam reflected from the XOU.}
	\label{fig:arm2}
\end{figure}
	
Even if monochromating crystals are quite small, they still requires translation and rotation systems to ensure arc-second accuracy alignment. The beam expander with its alignment system will be mounted close to the monochromators, in the same chamber, in order to replaced at the same time when the energy has to be changed.
	
\subsubsection{Long arm, test chamber and 12 m-long tube \label{sec:arm2}}
The beam exits from the beam expander unit to illuminate the entrance pupil of the XOU under test (vertical cylindrical chamber on the left of Fig.~\ref{fig:arm1}). The XOU chamber has 2 valves in order to exchange the module without breaking the vacuum both in the two arms and reducing the pumping time.

The beam is finally focused at about 12~m and reflected toward the basement with deviation angle depending on the position in the petal of the telescope optics. The maximum beam deviation corresponds to the radius of the ATHENA optics. For this reason, the long arm is mounted in a special self-sustaining structure in which the tube can be tilted (Fig.~\ref{fig:arm2}) about an horizontal axis passing by center of the XOU chamber. To this end, the tube is appropriately counterbalanced. Because the vertical travel range of the tilted tube is about 2 meters, a groove in the basement is to be made: this will allow us keeping the components of \Beatrix at a convenient height for handling.

\section{Expected performances} \label{sec:simulazioni}
In this section we display two simulations with a 4 reflection monochromator with and without misalignment.
The simulation have been carry out with SHADOWvui, extension of XOP~2.4\cite{XOP2004}.

\subsection{Simulation with misalignment of 3.5~arcsec in the first crystal of the monochromator \label{sec:4diffractions-simulation}}

\begin{figure}[ht]
\centering
\includegraphics[width=0.95\linewidth]{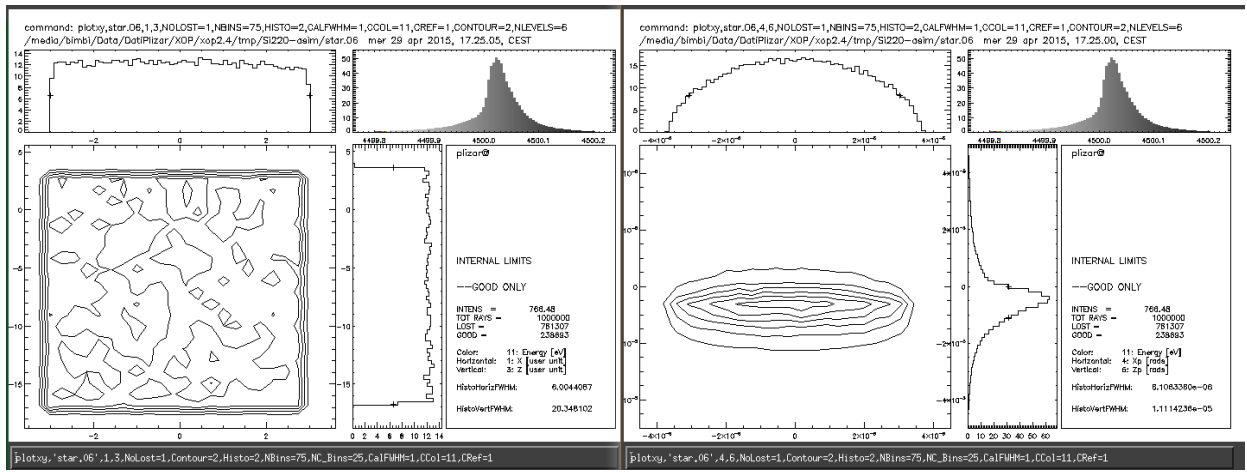}
\caption{Simulation with paraboloid, 4 perfect crystals Si(220), entrance crystal with 3.6~arcsec of misalignment and asymmetric cut Si(220). The expected beam divergence is approx. 2~arcsec of FWHM, corresponding to 1.5~arcsec HEW.}
\label{fig:shadow-4dif}
\end{figure}

Figure~\ref{fig:shadow-4dif} reports the expected performances of the setup previously described. In this simulation we used a monochromator with 4 crystals and a misalignment of 3.6~arcsec between the first and the second crystal.
The right window of Fig.~\ref{fig:shadow-4dif} gives the divergence of the beam in radians, while on the left we report the beam size in centimeters. The final beam has a maximum divergence of 2.3~arcsec FWHM and a size of 200~mm $\times$ 60~mm, as required. The expected flux on the XOU is 20 ph/sec/cm$^2$, assuming the X-ray source flux reported in Sect.~\ref{sec:source}.

\subsection{Simulation without misalignment \label{sec:no-disallineamento}}

\begin{figure}[ht]
	\centering
	\includegraphics[width=0.95\linewidth]{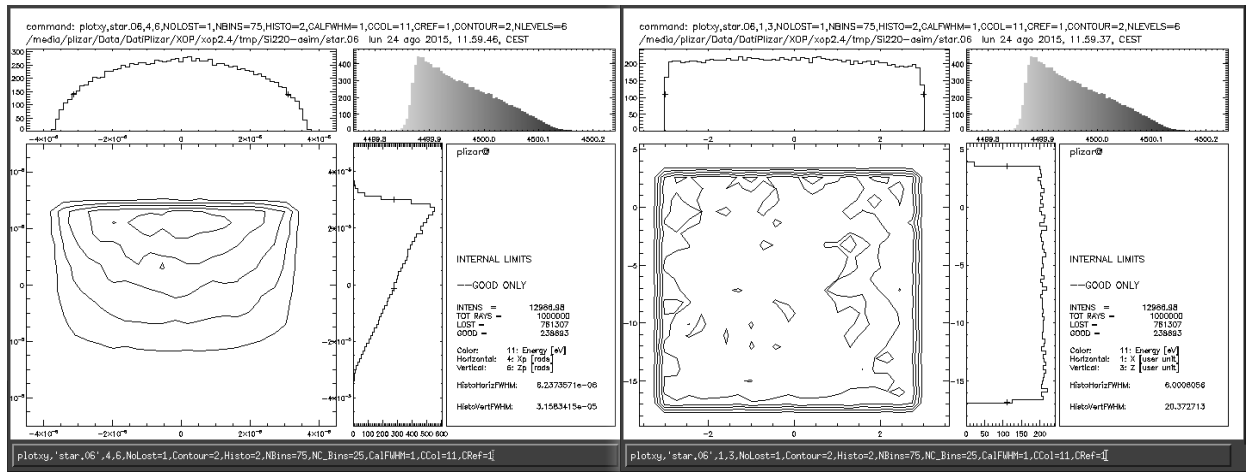}
	\caption{Simulation with paraboloid, 4 perfect crystals Si(220) and asymmetric cut Si(220). Crystals are perfectly aligned. Final divergence is 3 times larger than the specified value.}
	\label{fig:sim-Si220-nodisall}
\end{figure}

Fig.~\ref{fig:sim-Si220-nodisall} reports the expected spatial and angular distribution of the beam with a perfect alignment of the crystals of the monochromator. The divergence increases to 9-10~arcsec, while the photon flux has become 10 times higher than in the previous case. Hence, the present configuration can be used during the alignment phase: in the operative phase, the misalignment in the first crystal of the monochromator can be introduced to shrink the divergence down to the specified value.

\section{\label{sec:conclusioni} Conclusions}

BEaTriX will provide the capability to perform a direct, non-destructive characterization of single and stacked integrated mirrors developed for ATHENA in soft X-rays. The beam produced will be broad, monochromatic, parallel, collimated, and polarized in the incidence plane of the optic under test. The facility, to be realized at INAF/OAB, will be compact, fast to operate for a characterization in situ and in real time. Owing to its compactness, it will fit a small lab, and it will also be possible to reproduce at the industrial site of XOU production. The focal length will be easy to adapt removing or adding segments of the long tube. It will also enable the alignment of the parabolic/hyperbolic stacks under X-rays.

Thanks to the grant awarded to the AHEAD (Activities in the High Energy Astrophysics Domain) project by the European Commission, the operative phase of \Beatrix starts in September 2015.

\acknowledgments
We thank Manuel Sanchez Del Rio (ESRF, Grenoble, France), Finn E. Christensen (DTU, Denmark), Marco
Barbera (Universit\`a di Palermo, Italy), Marcos Bavdaz (ESA/ESTEC, The Netherlands) for useful discussions.

\bibliographystyle{spiebib}

\end{document}